\documentclass[12pt,a4paper]{scrartcl}

\usepackage{graphicx}

\title{\Large Reality of Manned Flying Reactor}
\author{Dmitry Zhuridov\footnote{Email: dmitry.zhuridov@wayne.edu}
\\ \normalsize\itshape
Department of Physics and Astronomy, Wayne State University, Detroit, MI 48201, USA}

\begin{document}
\maketitle

\begin{abstract}
	\vspace{-2em}
	\small\noindent	  New concept for reducing dose radiation exposure, which helps to decrease the duration and cost of deep space human missions is introduced.  This concept can be efficiently realized, using modern materials, such as carbon nanotube composites.
\end{abstract}

\vspace{2em}
  
The dream of a flying reactor, which takes advantage of a nuclear power to bring humans to other planets, has been around the corner for many years~\cite{Bowles_Arrighi}. To develop such a spacecraft, NASA and other space agencies have established number of projects such as NERVA~\cite{NERVA}, Project Prometheus~\cite{Prometheus}, etc. However they were closed due to serious budget and technical problems. 

One of the main problems is to shield the crew from the radiation of the onboard reactor. Traditional shielding was so thick and heavy that it would significantly complicate liftoff~\cite{Bowles_Arrighi}. The current NASA Advanced Exploration System project with Nuclear Cryogenic Propulsion Stage (NCPS)~\cite{NCPS} has same challenge. At present, the mass of required traditional shielding exceeds 7 tones~\cite{shielding_mass}.
However this problem can be easily solved by making a spacecraft from two separate modules:  nuclear booster and manned module, which are connected to each other by a long rope (several ropes for safety reasons) with length of order 1~km, as shown in Fig.~1, where $\alpha$ is the small angle between the rope line and the directions of jetstreams. 
Both modules should have a number of small orbit correction engines for making mutual maneuvers.
\begin{figure}[tb]
\begin{center}
     \includegraphics[width=1\textwidth]{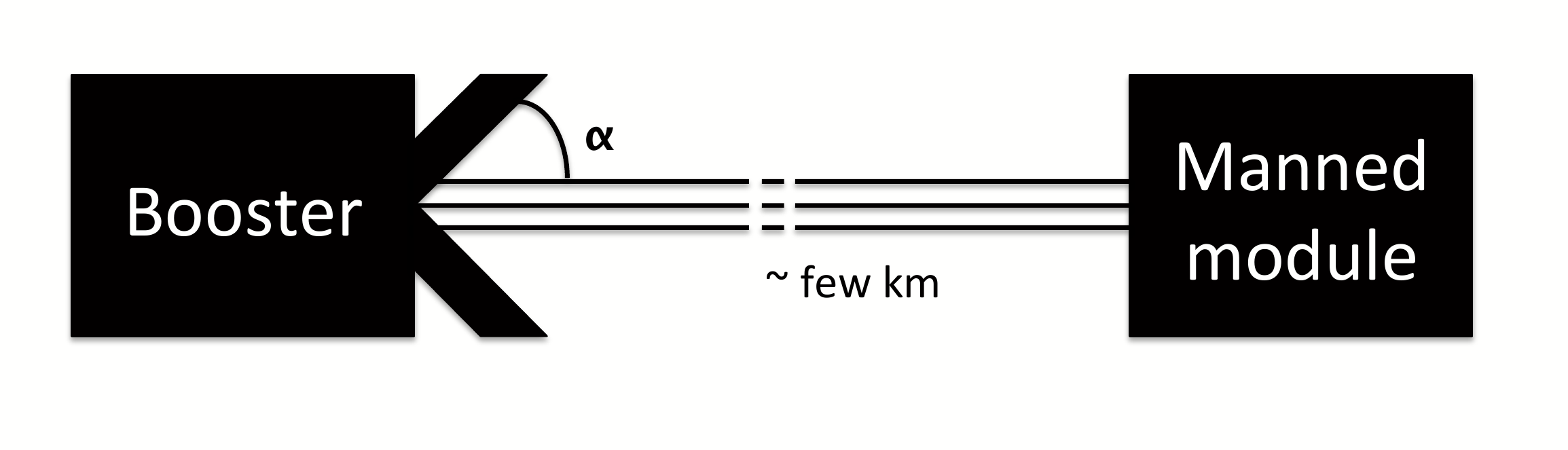}
\caption{Scheme of a spacecraft.  (Angle $\alpha$ is exaggerated for clarity.)}
\label{default}
\end{center}
\end{figure}

The required rope(s) can be made of modern developing materials, such as carbon nanotube (CNT) composites, to reduce the mass of the spacecraft. A novel-processing approach to production of CNTs that can be easily scaled up for industrial production, was recently introduced in Ref.~\cite{CNTcomposites}. In particular, the CNTs were stretched by pulling onto the rotating spool, which improves the tensile strength of the CNT composite ÒribbonÓ by approximately 90 percent (to an average of 3.5 gigapascals). The density of this stretch-wound composite is only 1.25 g$\cdot$cm$^{-3}$. Hence the mass of ropes is vanishing comparing to the rest of mass of the spacecraft.
Usage of a nuclear engine without a heavy traditional shielding will significantly decrease the time of a spaceflight to Mars and other destinations, as well as the dose radiation exposure from cosmic rays.
As additional bonus, the discussed ropes can be used for making the artificial gravity by rotation, when the cruising speed is achieved.
Moreover after significant coronal mass ejections (CME) from the sun in the direction to the spacecraft, the booster can be temporally placed between the sun and the manned module and used as additional shield from solar radiation. Hence the booster should be properly designed in a way, which allows to effectively use it together with the shield in the manned module in a cascade shielding system against CME.

Naively, possible breaking of a rope during the spaceflight is a danger and a minor point of the proposed scheme. However, if one or several ropes will be broken in any incident, such as a meteoroid impact, they can be replaced by the reserved ropes (by using proper simple devices).


Note that the proposed configuration of the spacecraft in Fig.~1 recalls the design project in Ref.~\cite{Mendell}. However the huge aluminum square truss is replaced by lightweight CNT ropes, and mass of the shielding from the radiation of the onboard reactor is significantly decreased due to much larger distance between the booster and the manned module. Moreover the angle $\alpha$ is significantly decreased for the same reason, which essentially increases the spacecraft's propulsion. 

In conclusion, we introduced a new scheme of a manned spacecraft with nuclear propulsion, which can be effectively implemented using modern materials, such as CNT composites.  This concept allows to significantly reduce mass of the radiation shielding and the spacecraft itself. Hence the duration of human missions to other planets, the crew health risk, and eventually the total mission cost can be essentially lowered.

\end{document}